\documentclass[11pt]{article}
\usepackage{graphicx}

\newcommand{\be}{\begin{equation}}
\newcommand{\ee}{\end{equation}}
\newcommand{\bea}{\begin{eqnarray*}}
\newcommand{\eea}{\end{eqnarray*}}
\newcommand{\ba}{\begin{eqnarray}}
\newcommand{\ea}{\end{eqnarray}}

\newcommand{\al}{\alpha}
\newcommand{\om}{\omega}
\newcommand{\HO}{{\cal O}}
\newcommand{\vp}{\varphi}

\newcommand{\ga}{\gamma}

\newcommand{\dt}{\delta}

\newcommand{\sg}{\sigma}
\newcommand{\lb}{\lambda}

\renewcommand{\baselinestretch}{2}
\begin{document}
\begin{flushright} UCL-IPT-99-08
\end{flushright}

\vspace*{5mm}

\begin{center}
\Large{\bf The observable light deflection angle}
\end{center}

\vspace*{15mm}

\begin{center}
\Large{\it J.-M. G\'erard and S. Pireaux }
\end{center}

\vspace*{5mm}

\begin{center}
Institut de Physique Th\'eorique\\
Universit\'e catholique de Louvain \\
 B-1348 Louvain-la-Neuve,
Belgium
\end{center}

\vspace*{10mm}
\begin{center}
\Large{To the memory of Jacques Demaret}
\end{center}

\thispagestyle{empty}

\vspace*{40mm}

\renewcommand{\baselinestretch}{1.4}

\begin{abstract}

The physical deflection angle of a light ray propagating in a space-time
supplied
with an asymptotically flat metric has to be expressed in terms of the impact
parameter.

\end{abstract}

\renewcommand{\baselinestretch}{2}

\newpage

\section{Introduction}

Light deflection by a gravitational source played a crucial role in the early
establishment of General Relativity. Nowadays, gravitational optics is
systematically
used to search for elusive dark matter candidates (such as white dwarfs)
through
microlensing effects \cite{1}.

>From a theoretical point of view, light deflection also provides us with a
rather
unique tool to test relativistic theories of gravity at various distance
scales.
In most of these theories, the exact value of the predicted deflection
angle requires
some numerical integration. Indeed, elliptic functions of the first kind
are already
necessary in the simple case of General Relativity. Therefore, approximate
analytical expressions are often used in the literature, both in the weak
and strong
field limits. In this short note, we would like to point out that only
expressions in terms of the physical impact parameter are meaningful.
Although this
statement is rather academic in the case of General Relativity, we
illustrate how it
might have quantitative consequences for extended theories of gravity.

\section{The deflection angle ...}

Let us consider a static and spherically symmetric space-time defined by
the line
element
\be
ds^2 = A^2 (r) c^2 dt^2 - B^2 (r) dr^2 - D^2 (r) r^2 (d\theta^2 + \sin^2
\theta d
\vp^2).
\ee
The null geodesics equations in the $\theta = \frac{\pi}{2}$ plane lead to the
following equation for the photon trajectories :
\be
\left( \frac{du}{d\vp} \right)^2 + \frac{D^2}{B^2} u^2 = \frac{1}{b^2}
\frac{D^4}{A^2B^2}
\ee
with
$$
u \equiv \frac{1}{r}.
$$
If space-time is asymptotically flat, $b$ is simply the impact parameter
which can be
expressed in terms of the radial coordinate of the point $(r_0,\vp_0)$ of
closest
approach
\be
b = \frac{D(r_0)}{A(r_0)} r_0.
\ee
Moreover, in such a space-time, the deflection angle $\hat \al$ is usually
defined by
\ba
\hat \al (r_0) &\equiv& 2 \int^{\vp_0}_0 d \vp - \pi \nonumber \\
&=& 2 \int^\infty_{r_0} \frac{B(r)}{D(r)} \biggl\{ \Bigl( \frac{r}{r_0}\Bigr)^2
\frac{D^2(r)}{D^2(r_0)} \frac{A^2(r_0)}{A^2 (r)} - 1
\biggr\}^{-\frac{1}{2}} \frac{dr}{r} - \pi
\ea
and requires, in general, a numerical integration.

{\bf In the weak field approximation}, the standard Eddington-Robertson
parametrization is defined by
\ba
A^2 (\rho) &\simeq& 1 + 2 \al \left( \frac{V}{c^2} \right) + 2 \beta \left(
\frac{V}{c^2} \right)^2 + \frac{3}{2} \xi \left( \frac{V}{c^2} \right)^3
\nonumber
\\
B^2 (\rho) &=& D^2 (\rho) \nonumber \\
&\simeq & 1 - 2 \ga \left( \frac{V}{c^2} \right) + \frac{3}{2}
\dt \left(\frac{V}{c^2}\right)^2
\ea
with
$$
V(\rho) \equiv - \frac{GM}{\rho},
$$
the gravitational potential expressed in terms of the isotropic radial
coordinate
$\rho$. In this approximation,  the deflection angle as a function of the
closest
approach distance reads
\be
\hat \al_\om (\rho_0) =
\frac{4GM}{c^2 \rho_0}
\biggl\{
\Bigl( \frac{\al + \ga}{2}   \Bigr) +
\Bigl[- \frac{(\al + \ga)^2}{2} +  \frac{(8\al^2 + 8 \al \ga - 4 \beta + 3
\dt)}{16}
\pi   \Bigr] \frac{GM}{c^2\rho_0}
 \biggr\} + \HO \Bigl( \frac{V^3}{c^6} \Bigr).
\ee
In particular, one easily proves the absence of light deflection in a
space-time
supplied with a conformally flat metric
\be
g_{\mu\nu} = \left(1+\frac{V}{c^2}\right)^2 \eta_{\mu\nu}.
\ee

\section{... in General Relativity ...}

The Schwarzschild coordinates are more appropriate to study light
deflection in the
vicinity of a strong gravitational field. In the case of General
Relativity, they
correspond to the choice
\ba
A^2 (r) &=& B^{-2} (r) = 1-\frac{2m}{r} \nonumber \\
D^2 (r) &=& 1
\ea
with
$$
m \equiv \frac{GM}{c^2}.
$$
Assuming
\be
r_0 \geq 3 m
\ee
to ensure deflection, we obtain from Eqs. (4) and (8) the angle as an {\bf
exact}
function of the closest distance of approach \cite{2}
\be
\hat \al (r_0) =
4 \left( \frac{r_0}{q}\right)^{\frac{1}{2}}
\biggl[ F  \Bigl(\frac{\pi}{2},k    \Bigr) - F(\sg_0,k) \biggr ] - \pi.
\ee
Here, the elliptic integral of the first kind
\be
F (\sg,k) \equiv \int^\sg_0 \frac{dx}{\{1-k^2 \sin^2 x \}^{\frac{1}{2}}}
\ee
is characterized by the amplitude $\sg$ with
\be
\sg_0 = \arcsin \left\{ \frac{q - r_0 + 2m}{q - r_0 + 6m}\right\}^{\frac{1}{2}}
\ee
and the modulus
\be
k = \left\{ \frac{q-r_0 + 6m}{2q}\right\}^{\frac{1}{2}}
\ee
where
$$
q \equiv \biggl\{ \Bigl(1-\frac{2m}{r_0} \Bigr) \Bigl(1+\frac{6m}{r_0} \Bigr)
\biggr\}^{\frac{1}{2}} r_0.
$$

{\bf In the weak field approximation}
\be
\epsilon \equiv \frac{m}{r_0} \ll 1
\ee
and the modulus is small
\be
k^2 = 4 \epsilon (1-3\epsilon) + \HO (\epsilon^3).
\ee
Expanding the elliptic integrals
\ba
F\left(\frac{\pi}{2} , k \right) &\simeq& \frac{\pi}{2} \left(1+\epsilon -
\frac{3}{4} \epsilon^2\right) \nonumber
\\
F\left(\sg_0 , k \right) &\simeq& \frac{\pi}{4} \left(1+\epsilon -
\frac{3}{4} \epsilon^2\right) - \epsilon
\ea
we obtain
\be
\hat \al_\om (r_0) \simeq
\frac{4GM}{c^2r_0}
\biggl\{ 1 + \Bigl[-1 + \frac{15}{16} \pi \Bigr] \frac{GM}{c^2r_0}  \biggr\}.
\ee
in the Schwarzschild coordinates. On the other hand, Eq. (6) implies
\be
\hat \al_\om (\rho_0) \simeq
\frac{4GM}{c^2\rho_0}
\biggl\{ 1 + \Bigl[-2 + \frac{15}{16} \pi \Bigr] \frac{GM}{c^2 \rho_0}
\biggr\}
\ee
in the isotropic coordinates, since the Eddington-Robertson parameters
defined in Eq.
(5) are conventionally normalized to unity for General Relativity.

The apparent second order discrepancy between Eqs. (17) and (18) can be
understood in
the following way. The deflection angle being by definition observable, it
has to be
fully expressed in terms of measurable, i.e. coordinate-independent,
quantities. Here,
the closest distance of approach ($r_0$ in Schwarzschild coordinates,
$\rho_0$ in
isotropic coordinates) is obviously not such a measurable quantity. Its
corresponding substitution by the impact parameter $b$ according to Eq. (3)
\ba
b &\simeq& r_0 + m \nonumber \\
  &\simeq& \rho_0 + 2m
\ea
reconciliates our results of calculations performed with two physically
equivalent
forms of the metric. In General Relativity, the observable deflection angle is
correctly given by
\be
\hat \al_\om (b) =
\frac{4GM}{c^2 b}
\biggl\{ 1 + \Bigl[0 + \frac{15 \pi}{16}\Bigr] \frac{GM}{c^2 b}  \biggr\} + \HO
\left( \frac{m^3}{b^3} \right).
\ee
This is the analog of what has been noted \cite{3} for the radar-echo
experiment where
the observable transit time has to be expressed in terms of the measurable
orbital
parameters (periods and eccentricities).

{\bf In the strong field approximation}
\be
\epsilon \equiv 1 - \frac{3m}{r_0} \ll 1
\ee
and the modulus is close to one
\be
k^2 = 1 - \frac{4}{3} \epsilon + \HO (\epsilon^2).
\ee
Expanding again the elliptic integrals
\ba
F \left( \frac{\pi}{2} , k\right) &\simeq& \ln \frac{4}{\sqrt{1-k^2}}
\nonumber\\
F\left( \arcsin \frac{1}{\sqrt{3}} , 1 \right) &\simeq& 0.65848,
\ea
we obtain a simple
analytical expression for the deflection angle of a light ray grazing a black
hole\cite{2} :
\be
\hat \al_s (b) \simeq \ln \left( \frac{3.482 m}{b - 3 \sqrt{3} m} \right).
\ee
The  Eqs. (20) and (24) provide us with useful (weak and strong field)
approximations to study lensing effects in terms of the observable
deflection angle
$\hat
\al (b)$. They readily replace the unpractical exact expression obtained
from Eq.
(10) after substitution of $r_0$ by $b$ according to Eqs. (3) and (8).
However, the
second  order contribution to the deflection angle in Eq. (20) only
improves the weak
field approximation in a  narrow domain of the impact parameter (see Fig.
1). In that
sense, our  comment on the ambiguous coordinate dependence of $\hat \al (r_0)$
defined in Eq. (4) is rather academic as far as General Relativity is
concerned.

\section{... and beyond}

Let us now consider the coupled Einstein-improved scalar action
\be
S = - \frac{1}{2 \kappa} \int d^4 x \sqrt{-g} R + \frac{1}{2} \int d^4 x
\sqrt{-g}
\left[ g^{\mu\nu}
\Phi_{,\mu} \Phi_{,\nu} + \frac{1}{6} R \Phi^2 \right]
\ee
with
$$
\kappa \equiv \frac{8 \pi G}{c^4}.
$$
The non-minimal coupling of the massless scalar $\Phi$ to the curvature
invariant
ensures the vanishing of the stress tensor's trace, as required by
conformal symmetry
\cite{4}.

It is however well-known \cite{5} that  an appropriate rescaling of the
metric
\be
\tilde g_{\mu\nu} = \left( 1 - \frac{\kappa}{6} \Phi^2 \right) g_{\mu\nu}
\ee
together with a redefinition of the scalar field
\be
\Phi = \left( \frac{6}{\kappa}\right)^{\frac{1}{2}} \tanh \biggl[  \Bigl(
\frac{\kappa}{6}
\Bigr)^{\frac{1}{2}} \widetilde \Phi
\biggr]
\ee
recast the theory into the minimal form
\be
\widetilde S =
- \frac{1}{2 \kappa} \int d^4 x \sqrt{- \tilde g} \widetilde R +
\frac{1}{2} \int d^4
x \sqrt{- \tilde g} \tilde g^{\mu\nu} \widetilde \Phi_{,\mu} \widetilde
\Phi_{,\nu}.
\ee
The most general static, spherically symmetric and asymptotically flat
exact solution
to the corresponding equations of motion
\ba
\widetilde R_{\mu\nu} &=& \kappa \  \widetilde \Phi_{,\mu} \widetilde
\Phi_{,\nu}
\nonumber\\
\makebox{\hfill$
\makebox[0mm]{\raisebox{0.3mm}[0mm][0mm]{\hspace*{5.1mm}$\sqcap$}}$
$
\sqcup$ }
\widetilde \Phi  &=& 0
\ea
is given by \cite{6}
\ba
A^2 (r) &=& B^{-2} (r) = \left( 1 - \frac{2m}{\lb r} \right)^\lb \nonumber\\
D^2 (r) &=& \left( 1 - \frac{2m}{\lb r} \right)^{1-\lb}
\ea
for the metric, and
\be
\widetilde \Phi = \left( \frac{1-\lb^2}{2\kappa}\right)^{\frac{1}{2}} \ln
\left( 1 -
\frac{2m}{\lb r} \right)
\ee
for the scalar field. If we assume a positive Newton constant $G$ (i.e.
$\kappa > 0)$,
then
\be
0 < \lb \leq 1
\ee
and we recover the standard Schwarzschild solution of General Relativity
when $\lb$
goes to one.

Light deflection requires now
\be
r_0 > \left( 2 + \frac{1}{\lb} \right) m.
\ee
{\bf In the weak field approximation},  Eqs.  (4) and (30) imply
\be
\hat \al_\om (r_0) \simeq \frac{4GM}{c^2r_0}
\Biggl\{ 1 + \biggl[ -2 + \frac{1}{\lb} + \Bigl( 1 - \frac{1}{16 \lb^2}
\Bigr)  \pi
\biggr]  \frac{GM}{c^2r_0}
\Biggr\}
\ee
On the other hand,  the Eddington-Robertson parameters of the minimal
Einstein-massless scalar theory  read
\ba
\al &=& 1 \nonumber\\
\beta &=& 1 \nonumber\\
\ga &=& 1 \nonumber\\
\dt &=& \frac{4}{3} \left( 1 - \frac{1}{4 \lb^2} \right)
\ea
such that
\be
\hat \al_\om (\rho_0) \simeq
\frac{4GM}{c^2 \rho_0}
\Biggl\{ 1 + \biggl[ -2 + \Bigl( 1 - \frac{1}{16 \lb^2} \Bigr) \pi \biggr]
\frac{GM}{c^2 \rho_0}
\Biggr\}.
\ee
Since a deviation from General Relativity only arises at the second order, the use ofthe physical 
impact parameter $b$ according to Eq. (3)
\ba 
b &\simeq&
r_0 + \left( 2 - \frac{1}{\lb} \right) m \nonumber\\
&\simeq& \rho_0 + 2m
\ea
is now crucial to obtain the observable deflection angle
\be
\hat \al_\om (b) =
\frac{4GM}{c^2 b}
\Biggl\{ 1 + \Bigl( 1 - \frac{1}{16 \lb^2} \Bigr) \pi
\frac{GM}{c^2 b }
\Biggr\} + \HO \left( \frac{m^3}{b^3}\right).
\ee
The same result is obtained by working
directly in the improved action basis characterized by
\ba
\al &=& 1 \nonumber\\
\beta &=& \frac{5}{6} + \frac{1}{6 \lb^2} \nonumber\\
\ga &=& 1 \nonumber\\
\dt &=& \frac{10}{9} - \frac{1}{9 \lb^2}.
\ea
The authors of ref. \cite{7} only gave the coordinate-dependent Eq. (4) to
analyse
the quantitative modifications of lensing characteristics in the presence
of the
massless scalar field $\widetilde \Phi$. We have just argued that Eqs. (4)
\underline{and} (3) should always be carefully handled to get rid of the
closest
distance of approach dependence. The deflection angle $\hat \al$ is then
correctly
expressed in terms of the physical impact parameter and does not depend on the
arbitrary choice of coordinate system.

\section{Conclusion}

Any observable has to be expressed in terms of measurable quantities which are
coordinate-independent. The use of the impact parameter is therefore
mandatory for
the deflection angle of a light ray propagating in a space-time supplied
with an
asymptotically flat metric. To our knowledge, this point has been
overlooked in the
literature.

\vspace*{5mm}

\par\noindent
{\Large{\bf Acknowledgements}}

We would like to thank Elisa Di Pietro and Luc Haine for discussions.

\newpage

\begin{figure}
		\begin{center}
  	\includegraphics[width=\linewidth]{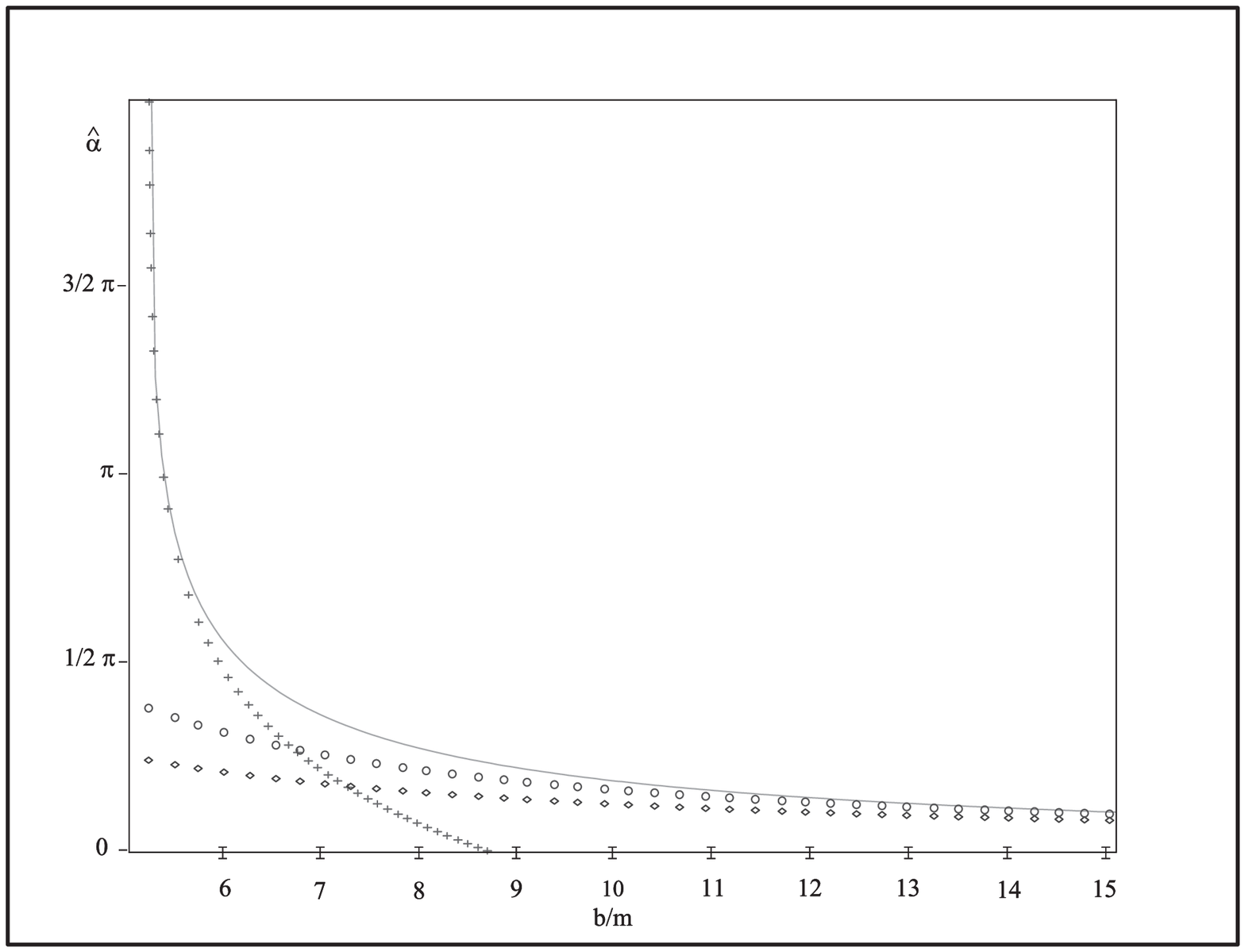}
  	\caption{{\footnotesize The light deflection angle $\hat \alpha$
predicted by
General Relativity, as a function of the impact parameter $b$. The smooth curve
corresponds to the exact expression based on Eq. (10) after substitution of the
closest distance of approach $r_0$ by $b$ according to Eqs. (3) and (8).
Crosses
stand for the strong field approximation given in Eq. (24). Diamonds and
circles
represent respectively the first and second order approximations in the
weak field
limit defined by Eq. (20).}}
		\end{center}
\end{figure}

\end{document}